\documentclass[letterpaper, 10pt, conference]{ieeeconf}
\IEEEoverridecommandlockouts
\usepackage{amssymb}
\usepackage{amsmath}

\usepackage{amsthm}
\usepackage{cite}
\usepackage{graphicx}
\usepackage{hyperref}
\usepackage{xcolor}

\theoremstyle{definition}
\newtheorem{definition}{Definition}

\title{\LARGE \bf Toward Neuronal Implementations of Delayed Optimal Control}

\author{Jing Shuang (Lisa) Li
	\thanks{J.S.L. is with the Department of Electrical Engineering and Computer Science at the University of Michigan, Ann Arbor. {\tt\small jslisali@umich.edu}.}
}

\begin{document}

\maketitle

\begin{abstract}
Animal sensorimotor behavior is frequently modeled using optimal controllers. However, it is unclear how the neural circuits within the animal's nervous system implement optimal controller-like behavior. In this work, we study the question of implementing a delayed linear quadratic regulator with linear dynamical ``neurons'' on a muscle model. We show that for any second-order controller, there are three minimal neural circuit configurations that implement the same controller. Furthermore, the firing rate characteristics of each circuit can vary drastically, even as the overall controller behavior is preserved. Along the way, we introduce concepts that bridge controller realizations to neural implementations that are compatible with known neuronal delay structures. 
\end{abstract}

\section{Introduction and motivation}

Control theory --- particularly, linear optimal control --- has emerged as the leading framework for sensorimotor modeling of various species and behaviors \cite{Li2004, Scott2004, Todorov2004, Franklin2011, Schultheis2021}. These works typically model input-output behaviors using a parameterized controller (e.g., linear quadratic regulator), and tune parameters (e.g., state and input penalty) to match data (e.g., movement trajectories, muscle forces) when available. Most models do not consider the underlying implementation of these input-output behaviors. Indeed, this is a common criticism of controls-based models --- they capture behavior without revealing the underlying implementation \cite{Scott2012, Loeb2012}. 

\begin{figure}[h]
  \centering
\includegraphics[width=\hsize, page=1]{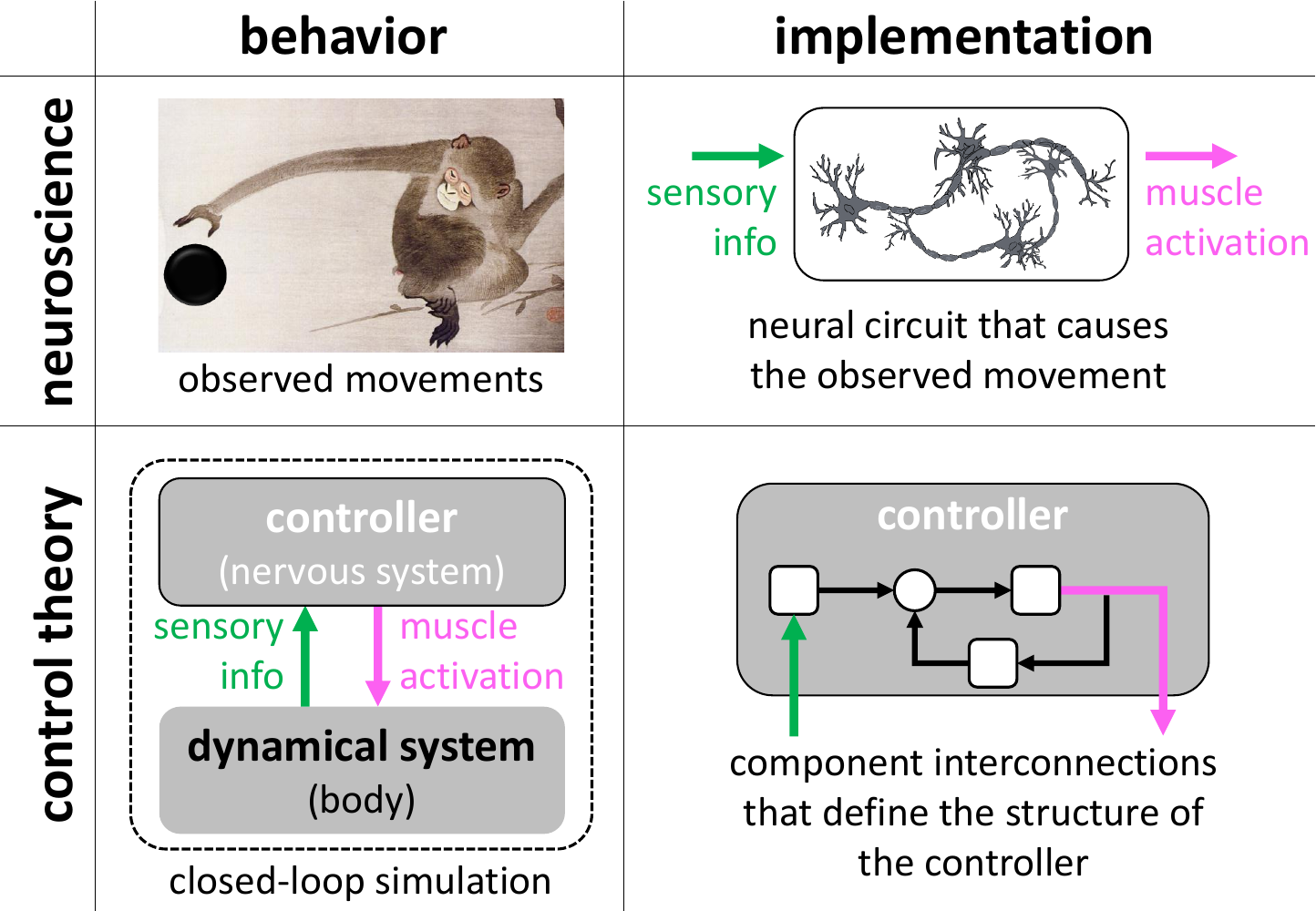}
  \caption{Behavior vs. implementation in neuroscience and control theory. While control theory offers explanations for behaviors seen in neuroscience experiments, it currently offers few explanations for neural circuitry (i.e., implementation) underlying these behaviors.} \label{fig:behavior_implementation}
\end{figure}

Current approaches to relate controller behavior to implementation roughly fall into two categories: anatomy-centered and learning-centered. In anatomy-centered approaches, neuroscientists associate parts of the controller with areas of the brain \cite{Miall2008, Dewolf2011, Franklin2011}. In learning-centered approaches, computer scientists create neurally-plausible approximations of controllers by training artificial neural networks on controller tasks \cite{Bekolay2014nengo, Goldsmith2020, Stagsted2020}. More recently, control theorists have joined the fray, drawing parallels between controller connectivity patterns and neuronal connectivity patterns \cite{Stenberg2022_IFP2, Li2022_IFP3, Li2023_PNAS}. All of these approaches have a common feature: they focus on one specific implementation of the controller. However, a given controller can be implemented in many different ways, similar to how a given transfer function has many potential state-space realizations. This is a longstanding observation in both control theory and neuroscience \cite{Krakauer2017}, but we do not yet have tools to rigorously explore the relationship between behavior and implementation in the neuroscience setting. The primary goal of this paper is to begin to address this research gap. Fig. \ref{fig:behavior_implementation} summarizes behavior and implementation in the context of neuroscience and control theory.

We focus on delayed optimal controllers\cite{Stenberg2022_IFP2, Li2023_PNAS}, which incorporate conduction delays of neurons (Section \ref{sec:optimal_delayed_controller}). We introduce necessary concepts that translate controller realizations to neural circuits (Section \ref{sec:compatibility}), and study different circuits associated with a given controller (Sections \ref{sec:microcircuits} and \ref{sec:simulations}). Throughout the paper, we ground our discussions in a simple neuromuscular example (Section \ref{sec:problem_setup}). However, the general methods presented extend to arbitrary linear time-invariant discrete-time systems. We conclude with some promising avenues of future investigation (Section \ref{sec:future_work}).

\section{Problem setup} \label{sec:problem_setup}
\begin{figure}[h]
  \centering
\includegraphics[width=\hsize, page=2]{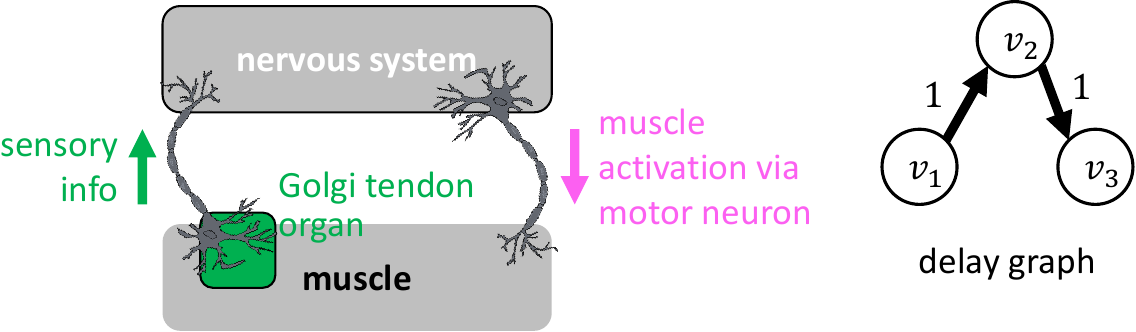}
  \caption{(Left) The neuromuscular control system; the muscle is the `plant', and the nervous system is the `controller'. The Golgi tendon organ senses the muscle force and communicates this to the nervous system, which fires a motor neuron to induce muscle activation. (Right) Delay graph associated with this system. There is a delay of 1 timestep from $v_1$ (Golgi tendon organ) to $v_2$ (nervous system), and a delay of 1 timestep from $v_2$ to $v_3$ (muscle). No other communication paths exist between the three vertices.} \label{fig:muscle_system}
\end{figure}

When using control theory to create models for neuroscience, we must first define what constitutes the controller, and what dynamical system this controller acts upon. Here, our controller is the nervous system, which acts upon a muscle subject to muscle dynamics. The nervous system receives information about the force of the muscle from the Golgi tendon organ, which is a sensor located on the muscle. The nervous system elicits force production from the muscle through muscle activation --- the muscle responds to the activity of the motor neuron that synapses onto it. The overall system is depicted in Fig. \ref{fig:muscle_system}. This framing allows us to compare controllers to neuronal circuits.

The nervous system is made of many neurons --- on the order of $10^{10}$ neurons for humans, and on the order of $10^{5}$ neurons for a fruit fly. The number of neurons involved in neuromuscular control depends on the motor task. Complex tasks involve many neurons in both the brain and spine; conversely, simple tasks, such as the one we study, are typically carried out in spinal circuits by relatively few neurons. We use a muscle model from \cite{Greene2024}: \footnote{Note: the model in \cite{Greene2024} includes two dimensionless parameters $D_0$ and $D_1$, which are randomized and represent different activation characteristics. In this work, we use $D_0 = 0$, $D_1 = 1$, which are within the biologically plausible range and allow for simple equations. The general findings in this paper do not change if we change these values, though the algebra will be slightly more complicated}
\begin{equation}
    \frac{df(t)}{dt} = \frac{f_\text{max}}{\tau (1 + e^{-r(t)})} - \frac{f(t)}{\tau}
\end{equation}
where $f(t)$ represents muscle force, and $r(t)$ represents the firing rate of the motor neuron; both are time-varying scalar functions. $f_\text{max}$ and $\tau$ are parameters representing the muscle's maximum force output and time constant. When the motor neuron is not firing, muscle force decays; when the motor neuron fires at a constant rate, muscle force rises to some fraction of the maximum muscle force. 

\begin{figure}[h]
\begin{minipage}{0.6\linewidth}
    \centering
    \includegraphics[width=0.95\hsize]{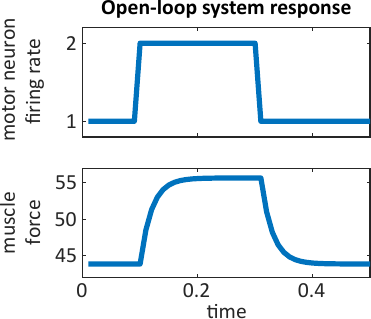}
\end{minipage}
\begin{minipage}{0.35\linewidth}
    \centering
    \caption{Open-loop system response of linear discrete-time neuromuscular model to a unit pulse input. Increasing motor neuron firing rate results in increased muscle force (with some transient dynamics); decreasing firing rate results in decreased force. Simulation parameters are given in Section \ref{sec:simulations}.} \label{fig:ol_sims}
\end{minipage}
\end{figure}

We convert this system to a linear discrete-time system. First, we linearize about equilibrium point $(\bar{f}, \bar{r})$, and define new state and input variables $\delta_f := f - \bar{f}$ and $\delta_r := r - \bar{r}$. Then, we discretize the system with sampling time $T_s$ to get:
\begin{equation} \label{eq:system}
    \delta_f(t+1) = A\delta_f(t) + B\delta_r(t)
\end{equation}
The input response of this system is shown in Fig. \ref{fig:ol_sims}.

In animals, sensing and actuation are subject to neuronal conduction delays. For humans, a sensory signal from a muscle on the foot must travel from the foot to the spine, and an actuation signal must travel from the spine back to the foot. Typical values are on the order of 10 ms each way \cite{More2018}. Smaller animals (e.g., fruit flies) have similar delay magnitudes \cite{Tuthill2016} --- though neural signals have a shorter distance to travel in smaller animals, they travel on smaller and therefore slower axons. To include delay in our formulation, we write our controller as:

\begin{equation} \label{eq:total_controller}
    \delta_r(t) = \mathcal{K}(\delta_f(0:t-T))
\end{equation}
where $\mathcal{K}$ is a linear map that depends on past values of its input, and $T$ is a non-negative integer denoting net sensorimotor delay (i.e., the sum of delays to and from the nervous system). In frequency-domain terms, let $\Delta_f(z) := \mathcal{Z}(\delta_f(t))$ and $\Delta_r(z) := \mathcal{Z}(\delta_r(t))$ where $\mathcal{Z}$ is the Z-transform. The transfer function $G(z) := \frac{\Delta_r(z)}{\Delta_f(z)}$ should have a relative degree of at least $T$, i.e., the difference between degree of the denominator and numerator should be no less than $T$.

\section{Optimal delayed controller} \label{sec:optimal_delayed_controller}

Given state and input penalty $Q$ and $R$, the LQ cost is
\begin{equation} \label{eq:cost}
    J = \sum_{t=0}^\infty \delta_f(t)^\top Q \delta_f(t) + \delta_r(t)^\top R \delta_r(t)
\end{equation}
We now seek a delayed controller of the form \eqref{eq:total_controller} that is LQ-optimal. We will use methods from previous work \cite{Stenberg2022_IFP2}. For simplicity, we assume that the system has been discretized so that there is one timestep of communication delay from controller to actuator, and one timestep from sensor to controller, i.e., net sensorimotor delay is $T=2$. Starting with system \eqref{eq:system}, we introduce two new variables:

\begin{equation} \label{eq:augmented_states}
    \gamma(t+1) = \mu(t), \quad \delta_r(t+1) = \gamma(t)
\end{equation}
$\mu$ represents \textit{intended} actuation, which is delayed by two timesteps before it reaches the physical system as $\delta_r$. $\gamma$ is a virtual signal representing the intended, delayed actuation. We define the \textit{augmented state} as $\chi(t) := \begin{bmatrix}  \delta_f(t)^\top & \gamma(t)^\top & \delta_r(t)^\top \end{bmatrix}^\top$, with augmented state equations
\begin{equation} \label{eq:augmented_system}
    \chi(t+1) = \begin{bmatrix}
        A & 0 & B \\
        0 & 0 & 0 \\
        0 & I & 0\\
    \end{bmatrix} \chi(t) + \begin{bmatrix}
        0 \\ I \\ 0
    \end{bmatrix} \mu(t)
\end{equation}

Equations for systems with more delays are given in \cite{Stenberg2022_IFP2}.

\begin{figure}[h]
\begin{minipage}{0.6\linewidth}
    \centering
    \includegraphics[width=0.95\hsize]{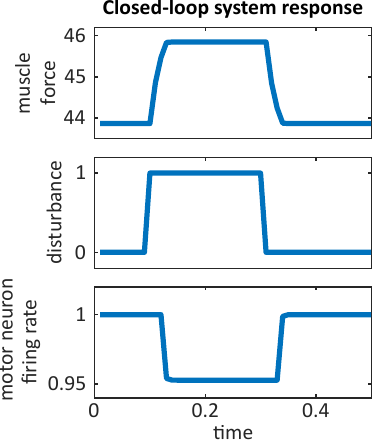}
\end{minipage}
\begin{minipage}{0.35\linewidth}
    \centering
     \caption{Closed-loop system response to a unit pulse disturbance. The control output (motor neuron firing rate) responds to the disturbance with a slight delay, and restores the force to equilibrium after the disturbance ends. Simulation parameters are given in Section \ref{sec:simulations}.} \label{fig:cl_sims}
\end{minipage}
\end{figure}

Cost \eqref{eq:cost} can be rewritten in terms of the augmented state:
\begin{equation} \label{eq:augmented_cost}
    J = \sum_{t=0}^\infty \chi(t)^\top \tilde{Q} \chi(t), \quad \tilde{Q} = \text{diag}(Q, 0, R)
\end{equation}
Minimizing \eqref{eq:augmented_cost} for system \eqref{eq:augmented_system} is a standard LQR problem; this is equivalent to minimizing \eqref{eq:cost} for original system \eqref{eq:system}, subject to delay constraints on the controller. The resulting optimal controller is $\mu(t) = K \chi(t)$. This can be rewritten as
\begin{equation} \label{eq:ifp_controller}
    \mu(t) = K_0 \delta_f(t) + K_1 \gamma(t) + K_2 \delta_r(t) \\    
\end{equation}
which, in conjunction with \eqref{eq:augmented_states}, make up the controller $\mathcal{K}$. We can convert these equations into z-domain to find the transfer matrix $G(z)$, which has a relative degree of 2:
\begin{equation} \label{eq:transfer_fn}
    G(z) := \frac{\Delta_r(z)}{\Delta_f(z)} = \frac{K_0}{z^2 - K_1z - K_2}
\end{equation}
Thus, the controller obeys the form of the delayed controller in \eqref{eq:total_controller} for $T=2$. The closed-loop system response to a pulse disturbance is shown in Fig. \ref{fig:cl_sims}. As expected, the controller responds with the appropriate amount of delay, and restores the force to equilibrium after the disturbance ends.

\begin{figure}[h]
\begin{minipage}{0.6\linewidth}
    \centering
    \includegraphics[width=0.83\hsize, page=3]{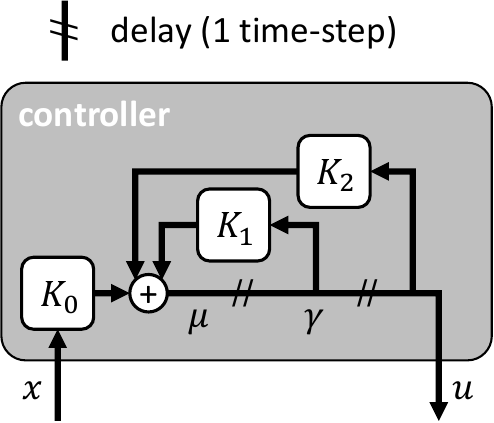}
\end{minipage} 
\begin{minipage}{0.35\linewidth}
    \centering
    \caption{Controller structure for the optimal delayed controller described by \eqref{eq:augmented_states} and \eqref{eq:ifp_controller}. Broken lines indicate one timestep of delay. This structure is not compatible with the specified communication delays.} \label{fig:ifp_diagram}
\end{minipage}
\end{figure}

Now, consider the resulting controller structure (Fig. \ref{fig:ifp_diagram}), which was studied in previous work \cite{Stenberg2022_IFP2, Li2023_PNAS}. This structure has a net delay of $T=2$, but is not compatible with the given problem. We specified one timestep of communication delay from sensor to controller, and from controller to actuator --- this should translate to delays at the input and output of the controller. However, in Fig. \ref{fig:ifp_diagram}, the delays are in the middle of the controller as opposed to at the input and output. This serves as a motivating problem: we are interested in studying controllers that are compatible (to be made precise in the next section) with a particular delay structure.

\textit{Remark:} The process shown in this section is one of three techniques for incorporating sensorimotor delays from previous work \cite{Stenberg2022_IFP2}. The two other techniques also suffer from the same issue of incompatibility.

\section{Compatibility of delayed controllers} \label{sec:compatibility}

\subsection{Realizations}
We now review the standard concept of controller realizations. Let a controller be represented in $z$-domain as $G(z)$. A \textit{realization} of a controller consists of matrices $(F, H, M, N)$ such that $G(z) = M(zI-F)^{-1}H + N$. The state-space controller associated with this realization is written as:
\begin{equation}
\begin{aligned}
x(t+1) &= F x(t) + H u(t) \\
y(t) &= M x(t) + N u(t)
\end{aligned}
\end{equation}
where we use $x, u$, and $y$ to indicate the internal state, input, and output of the \textit{controller} (as opposed to the plant). Every transfer function has infinite state-space realizations of varying size. If $(F, H, M, N)$ are a realization for $G(z)$, then so are similarity transforms of the realization, i.e., $(P^{-1}FP, P^{-1}H, MP, N)$, where $P$ is any invertible matrix. A \textit{minimal realization} is a realization in which all eigenvalues of $F$ are poles of $G(z)$; a \textit{non-minimal realization} is a realization in $F$ has more eigenvalues than $G(z)$ has poles.

\subsection{Structure, delay graphs, and compatibility}
We now introduce definitions of controller structures, delay graphs, and compatibility between the two.

\begin{definition}
    The \textit{structure} of a controller refers to the graphical representation of a specific realization of a controller. In this work, we will restrict graphical representations to only contain gain and addition blocks interconnected by scalar signals, possibly with delays.
\end{definition}

If two controllers share the same graphical representation (i.e., the configuration of blocks are the same, even if the gain values are different), we say that they have the same structure. Structure depends on the sparsity pattern of the realization matrices; while every controller has infinite state-space realizations, different realizations can translate to the same structure. A given controller can have multiple structures, and a given structure can be associated with multiple controllers.

\begin{definition}
    A \textit{delay graph} is a directed graph with a source $v_1$, sink $v_N$, and any number of intermediate vertices $v_i$, $i = 2 \ldots N-1$. Edge weights $E_{ij}$ represent the delay (in timesteps) from vertex $i$ to $j$. By convention, $E_{ii} = 0$. If vertex $i$ does not directly communicate to vertex $j$, $E_{ij}$ is the sum of weights for the shortest path from $i$ to $j$; if no such path exists, $E_{ij} = \infty$. The source $v_1$ represents the sensor that gives information to the controller; the sink $v_N$ represents the actuator that receives input from the controller. Intermediate vertices and edges represent additional details and constraints on the internal structure of the controller.
\end{definition}

The delay graph corresponding to the neuromuscular example is shown in the right side of Fig. \ref{fig:muscle_system}. This graph has $N=3$ and a corresponding edge weight matrix:
\begin{equation}
E = \begin{bmatrix}
0 & 1 & 2 \\
\infty & 0 & 1 \\
\infty & \infty & 0
\end{bmatrix}
\end{equation}

\begin{definition}
    For a given controller structure and delay graph, we can build a \textit{delay assignment} describing relationships between signals in the controller structure and vertices in the delay graph. Each signal must be assigned to exactly one vertex; the input to the controller ($u$) must be assigned to $v_1$, and the controller output ($y$) must be assigned to $v_N$. 
\end{definition}

Given a controller structure and delay graph, multiple delay assignments are possible.

\begin{definition}
    A delay assignment has an associated \textit{delay assignment matrix} $\tilde{E}$. $\tilde{E}_{ij}$ indicates the delay along the fastest path from a signal at vertex $i$ to a signal at vertex $j$. If no such signal exists, $\tilde{E}_{ij} = \infty$. By convention, $\tilde{E}_{ii} = 0$.
\end{definition}

\begin{definition}
    A controller structure is \textit{compatible} with a delay graph if there exists a delay assignment for this controller structure with an associated delay assignment matrix $\tilde{E}$ such that $\tilde{E}_{ij} \geq E_{ij} \quad \forall i,j$, where $E$ is the edge weight matrix of the delay graph.
\end{definition}

\begin{definition}
    A controller is \textit{compatible} with a delay graph if at least one realization of the controller has an associated controller structure that is compatible with the delay graph.
\end{definition}

\begin{figure}[h]
  \centering
\includegraphics[width=\hsize, page=8]{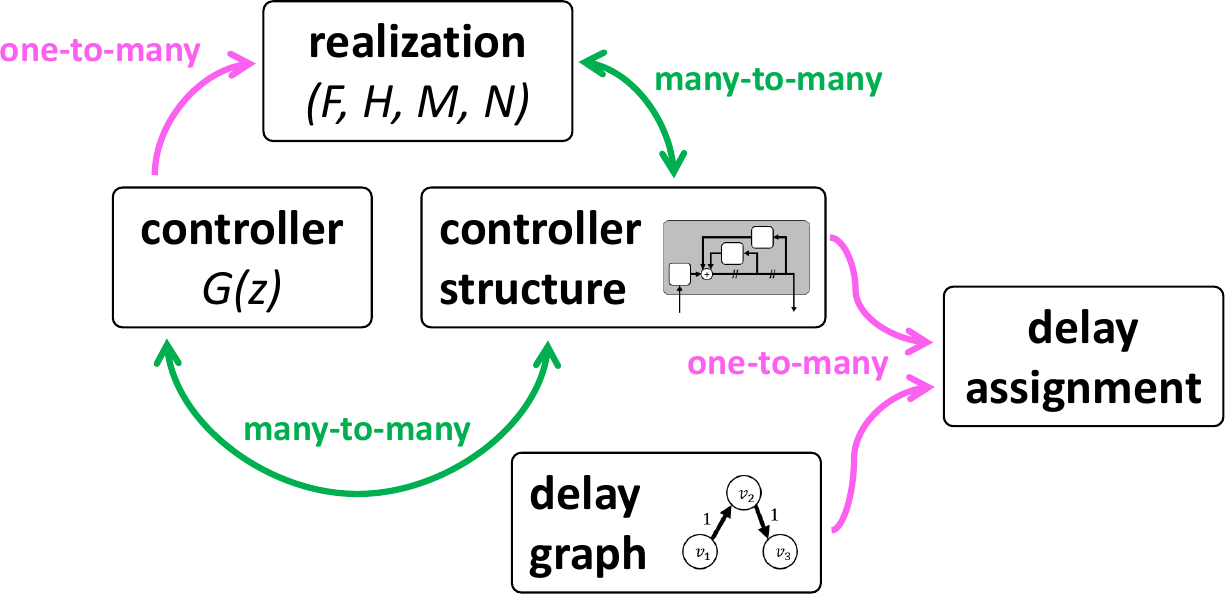}
  \caption{Relationships between controllers, realizations, controller structures, delay graphs, and delay assignments. Compatibility between a controller and a delay graph is assessed by comparing the controller's delay assignments and their associated matrices to the delay graph's edge weight matrix.} \label{fig:definitions}
\end{figure}

The goal of these (somewhat long) definitions is to be able to rigorously assess whether a given controller could be implemented on a given system with some specific pattern of delays. A summary of these definitions is given in Fig. \ref{fig:definitions}.

As an example, consider the controller structure from Fig. \ref{fig:ifp_diagram} and the delay graph in Fig. \ref{fig:muscle_system}. Let us try to build a compatible controller graph. Assign signal $x$ to $v_1$ and signal $u$ to $v_N$. The remaining signals to be assigned are $\gamma$ and $\mu$. Notice that there exists a signal path (through the $K_2$ block) from $u$ (at $v_3$) toward $\mu$ with no delay. If we assign $\mu$ to $v_2$, this would mean $\tilde{E}_{32} = 0$; if we assign $\mu$ to $v_1$, this would mean $\tilde{E}_{31} = 0$. In either case, compatibility is violated since $E_{31} = E_{32} = \infty$. The only option left is to assign $\mu$ to $v_3$, but we notice that there exists a delay-free signal path (through the $K_0$ and addition blocks) from $x$ (at $v_1$) to $\mu$ --- this would mean that $\tilde{E}_{13} = 0$, which is again incompatible since $E_{13} = 2$. Therefore, this controller structure is not compatible with this delay graph, since there exists no compatible delay assignment. This makes our statements at the end of Section \ref{sec:optimal_delayed_controller} mathematically precise.

\subsection{Creating a compatible delayed controller}

We now consider controller \eqref{eq:transfer_fn}, which has many potential realizations and structures. We showed above that one particular realization/structure is incompatible with the given delay graph. However, this does not preclude the possibility that other structures may be compatible --- indeed, several compatible structures exist, which we now explore.

To ensure compatibility with the delay graph, we first split $G(z)$ into $G(z) = G_3(z)G_2(z)G_1(z)$, and enforce that $G_3(z)$ and $G_1(z)$ have relative degree 1. These correspond to the delays along $E_{12}$ and $E_{23}$. Since $G(z)$ has relative degree 2, we see that the resulting $G_2(z)$ will have relative degree 0. Consider the simplest forms of $G_i(z)$:
\begin{equation}
\begin{aligned}
G_1(z) &= \frac{C_1}{z-\epsilon_1}, \quad G_3(z) = \frac{C_3}{z-\epsilon_3}\\
G_2(z) &= \frac{K_0}{C_1 C_3} \frac{z^2 - (\epsilon_1 + \epsilon_3)z + \epsilon_1 \epsilon_3}{z^2 - K_1 z - K_2}
\end{aligned} 
\end{equation}
for some constants $C_1, C_3, \epsilon_1, \epsilon_3 \in \mathbb{R}$, $C_1 \neq 0$, $C_3 \neq 0$.

The realization for $G_i(z)$, $i = 1, 3$ is 
\begin{equation}
\begin{aligned}
    x(t+1) &= \epsilon_i x(t) + C_i u(t) \\
    y(t) &= x(t)
\end{aligned}    
\end{equation}

\begin{figure}[h]
  \centering
\includegraphics[width=\hsize, page=4]{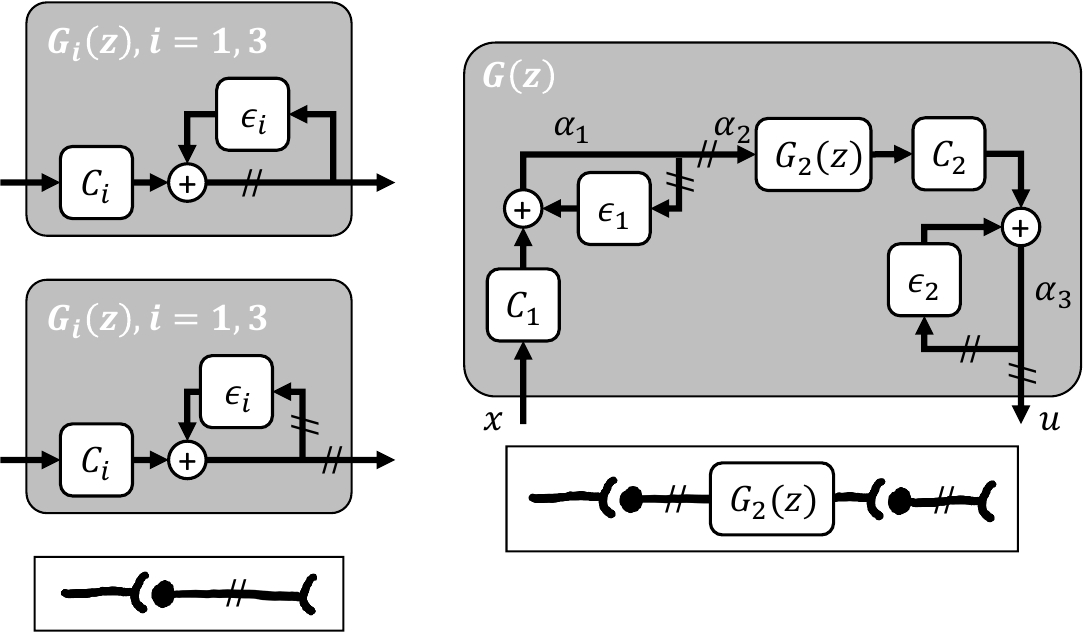}
  \caption{(Left) Two controller structures associated with $G_1(z)$ and $G_3(z)$. We use the structure on the bottom. (Right) Structure associated with $G(z)$. We also include the neural circuit interpretation of each structure.} \label{fig:g1g3}
\end{figure}

This is shown in Fig. \ref{fig:g1g3}. This is compatible if we assign $x$, $\alpha_1$ to $v_1$, $u$ to $v_3$, and $\alpha_2, \alpha_3, \alpha_4$, as well as any internal variables in $G_2(z)$ to $v_2$. Notice that even for this extremely simple first-order system, there are multiple possible structures (see Fig. \ref{fig:g1g3}). We choose the structure that, when combined with $G_2(z)$, produces a compatible structure.

The delay assignment matrix is:
\begin{equation}
    \tilde{E} = \begin{bmatrix}
        0 & 1 & \infty \\
        \infty & 0 & 1 \\
        \infty & \infty & 0
    \end{bmatrix}
\end{equation}
where $\tilde{E}_{12}$ was determined by the signal from $\alpha_1$ to $\alpha_2$, and $\tilde{E}_{23}$ was determined by the path from $\alpha_4$ to $u$. 

The general realization for $G_2(z)$ is
\begin{equation} \label{eq:general_realization}
\begin{aligned} 
    x_1(t+1) &= F_{11}x_1(t) + F_{12}x_2(t) + H_1 u(t) \\
    x_2(t+1) &= F_{21}x_1(t) + F_{22}x_2(t) + H_2 u(t) \\
    y(t) &= M_1 x_1(t) + M_2 x_2(t) + N u(t)
\end{aligned}
\end{equation}

By rewriting $G_2(z)$ as
\begin{equation}
    \frac{K_0}{C_1 C_3} + \frac{K_0}{C_1 C_3} \frac{(K_1 - \epsilon_1 - \epsilon_3) z + K_2 + \epsilon_1 \epsilon_3}{z^2 - K_1 z - K_2}
\end{equation}
we see that $N = \frac{K_0}{C_1 C_3}$ for every realization.

We show the resulting structure in the left panel of Fig. \ref{fig:full_diagram}. This structure exhibits some symmetry between the upper and lower portions (e.g., the paths through $H_1$ and $M_1$ vs. the paths through $H_2$ and $M_2$). Realizations with zeros will result in slight changes to the structure --- for instance, the controllable canonical realization:
\begin{equation}
\begin{aligned}
    F = \begin{bmatrix} 0 & 1 \\ K_2 & K_1\end{bmatrix}, \quad
    H = \begin{bmatrix} 0 \\ 1 \end{bmatrix} \\
    M = \frac{K_0}{C_1 C_3}\begin{bmatrix} K_2 +\epsilon_1 \epsilon_3 & K_1 - \epsilon_1 - \epsilon_3 \end{bmatrix}
\end{aligned}
\end{equation}
The resulting controller structure is shown in the left panel of Fig. \ref{fig:ccr_diagram}. We also have the observable canonical realization:
\begin{equation}
\begin{aligned}
    F = \begin{bmatrix} 0 & K_2 \\ 1 & K_1\end{bmatrix}, \quad M = \begin{bmatrix} 0 & 1\end{bmatrix} \\
    H = \frac{K_0}{C_1 C_3} \begin{bmatrix} K_2 +\epsilon_1 \epsilon_3 \\ K_1 - \epsilon_1 - \epsilon_3 \end{bmatrix}
\end{aligned}
\end{equation}
The resulting controller structure is shown in the left panel of Fig. \ref{fig:ocr_diagram}. As previously mentioned, controller structure depends on the sparsity pattern in the realization matrices. To be precise, different sparsity patterns are necessary but not sufficient for different structures. For instance, if we switch the order of $x_1$ and $x_2$ in the observable canonical realization, the resulting realization matrices would have different sparsity patterns, but the same structure.

\section{Neurons and microcircuits} \label{sec:microcircuits}

\begin{figure}[h]
  \centering
\includegraphics[width=0.7\hsize, page=9]{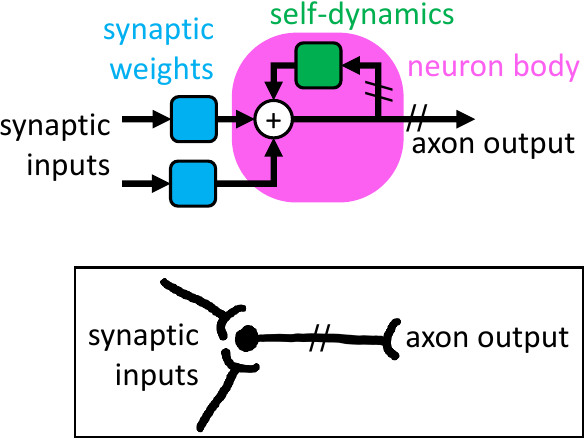}
  \caption{A stylized neuron, depicted as a controller structure (top) and equivalent neural circuit (bottom). Squares in the top diagram represent scalar gains. The synaptic inputs are multiplied by the synaptic weights, and added to the self-dynamics to produce the axon output (with some delay). The self-dynamics and computation occur in the ``neuron body'', which is depicted by the circle in the lower diagram.
  } \label{fig:neural_circuit}
\end{figure}

It is unintuitive to compare controller structures with circuits of neurons. We now introduce a stylized ``neuron'', which receives synaptic inputs (typically from the axons of other neurons) and produces an output on its own axon. The output of the neuron is a weighted combination of synaptic inputs and past outputs. We say that the neuron contains \textit{self-dynamics} because its output uses information from past outputs. Values of synaptic inputs and axon outputs represent firing rates. We can now use this stylized definition of a neuron to convert controller structural diagrams into ``neural'' circuits --- an example is given in Fig. \ref{fig:neural_circuit}. An axon can split into multiple branches (see Fig. \ref{fig:ccr_diagram}), allowing the output of the neuron to reach multiple downstream neurons. This is consistent with axon branching in neuroanatomy.

Our stylized neuron is similar to artificial neurons used in machine learning, with two key differences. Firstly, the self-dynamics in our neuron are conceptually reminiscent of a recurrent unit; secondly, our neuron lacks a nonlinear activation function. By avoiding this activation function, we are able to analytically and exactly translate controller structures to stylized neural circuits --- we defer exploration of more complicated and realistic neurons to future work.

For each realization considered in the previous section, we draw the corresponding neural circuit for $G_2(z)$ --- see Fig. \ref{fig:full_diagram} for the full realization, Fig. \ref{fig:ccr_diagram} for the controllable canonical realization, and Fig. \ref{fig:ocr_diagram} for the observable canonical realization. Note that the full circuit for $G(z)$ is simply the circuit for $G_2(z)$ plus an two extra neurons at the input and output (see bottom of Fig. \ref{fig:g1g3}). We can further simplify these circuits if we choose $\epsilon_i$ such that $K_2 + \epsilon_1 \epsilon_3 = 0$ or $K_1 - \epsilon_1 - \epsilon_3 = 0$ --- this results in circuits that look like the controllable canonical realization (Fig. \ref{fig:ccr_diagram}), but with the middle neuron having only one axon branch instead of two. 

Overall, we have three very distinct circuits, all of which --- in conjunction with $G_1(z)$ and $G_3(z)$ --- are compatible with the neuromuscular delay graph. Interestingly, although we confined ourselves to minimal realizations, these minimal realizations translate to circuits with different numbers of neurons. Additionally, even though the controllable and observable canonical realizations have the same number of zeros in their matrices, they have circuits with different numbers of neurons (2 and 3, respectively).

These three circuits --- in conjunction with $G_1(z)$ and $G_3(z)$ --- can implement \textit{any} second-order controller. This follows naturally from the generality of the three types of realizations we considered. This has interesting implications for neuronal circuits in the body --- namely, that a given circuit can capture a wide array of behaviors depending on the specific values of the synaptic weights and self-dynamics. This is reminiscent of the idea of canonical microcircuits within the brain \cite{nelson2002cortical}, in which the same circuit is found throughout the cortex (i.e., surface layer of the brain), and is hypothesized to perform some kind of universal computation. In our case, the ``universal computation'' associated with our circuits can be any second-order dynamical computation.

\textit{Remark:} It is possible to rework some methods from  \cite{Stenberg2022_IFP2} to make the resulting controller structure compatible with some sensorimotor delay graph, as is done in \cite{Karashchuk_2024}. However, the resulting structure is highly non-minimal --- indeed, for a scalar system, its dynamics can also be captured by one of the three circuits mentioned above.

\section{Simulations} \label{sec:simulations}

We perform numerical simulations to observe patterns of firing rates in the three circuits. We use the muscle model with parameters $\tau = 0.02$ seconds and $f_{\text{max}} = 60$ N, taken from \cite{Greene2024}. We linearize about $\bar{f} = 43.9$ N and $\bar{r} = 1$ spike/s, which represent values associated with some body equilibrium state\footnote{Even when the body is at equilibrium, most muscle forces are not at zero; this is associated with some nonzero firing rate from the motor neuron. For instance, when standing, our leg muscles must maintain some constant force output, without which we would fall down}. Let the one-way conduction delay between muscle and nervous system be be $10$ ms; we use step size of $10$ ms. The resulting linear discrete-time system is described by \eqref{eq:system} with $A = 0.61$ and $B = 4.64$. The net sensorimotor delay is $T = 2$ timesteps; the delay graph associated with the system is shown in Fig. \ref{fig:muscle_system}.

\begin{figure*}[h]
  \centering
\includegraphics[width=\hsize, page=7]{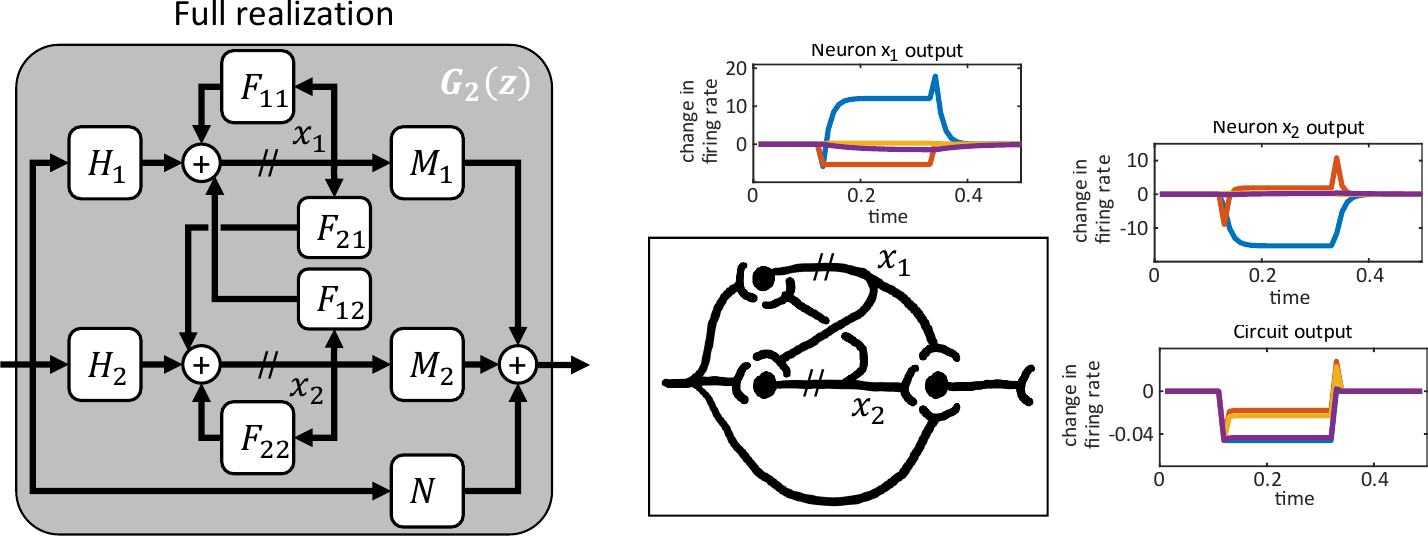}
  \caption{(Left) Controller structure for $G_2(z)$ for a general realization. (Right) Neural circuit associated with this controller structure. Time-series of changes in firing rate (relative to equilibrium) are shown for four different simulations corresponding to different randomized realization matrices. To generate these matrices, we first randomly generate $\epsilon_1$ and $\epsilon_3$, then compute the optimal realization using \eqref{eq:general_realization}. Then, we randomly generate a full-rank matrix and compute the similarity transform, which gives the final gain values used in the plots. \vspace{1em}} \label{fig:full_diagram}
\end{figure*}

\begin{figure*}[h]
  \centering
\includegraphics[width=\hsize, page=5]{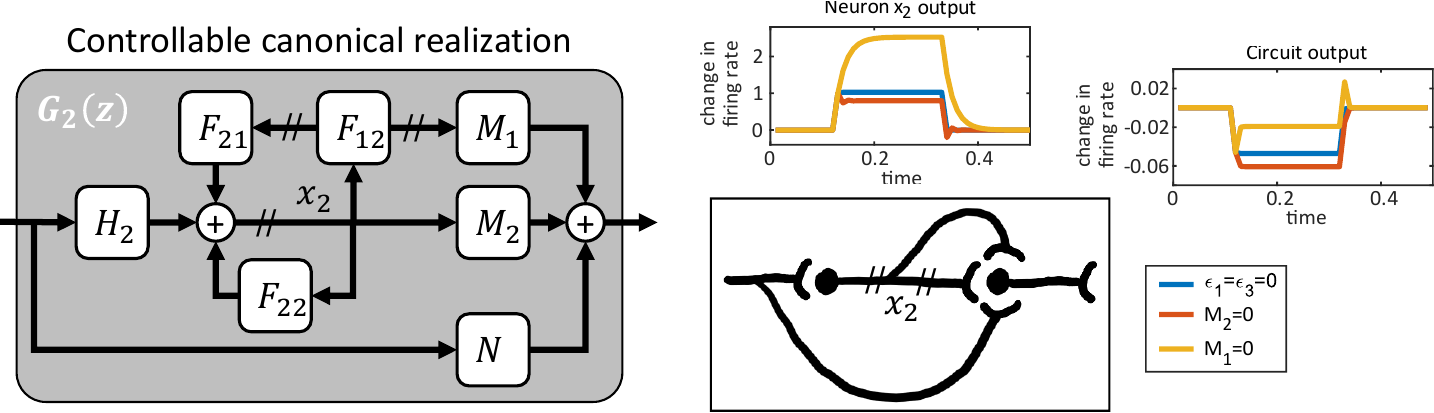}
  \caption{(Left) Controller structure for $G_2(z)$ for the controllable canonical realization. (Right) Neural circuit associated with this controller structure. Time-series of changes in firing rate (relative to equilibrium) are shown for three different simulations corresponding to different realization matrices. \vspace{1em}} \label{fig:ccr_diagram}
\end{figure*}

\begin{figure*}[h]
  \centering
\includegraphics[width=\hsize, page=6]{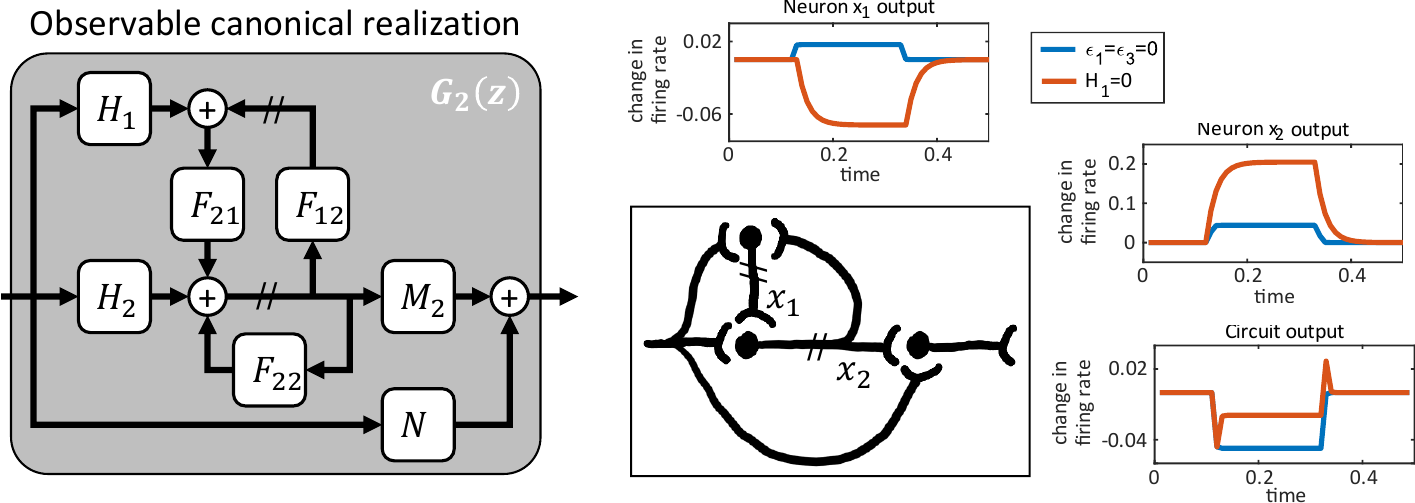}
  \caption{(Left) Controller structure for $G_2(z)$ for the observable canonical realization. (Right) Neural circuit associated with this controller structure. Time-series of changes in firing rate (relative to equilibrium) are shown for two different simulations corresponding to different realization matrices.} \label{fig:ocr_diagram}
\end{figure*}

For each circuit, we plot the axon outputs of each neuron in $G_2(z)$ in response to a pulse disturbance (Figs. \ref{fig:full_diagram}, \ref{fig:ccr_diagram}, \ref{fig:ocr_diagram}). We include simulations with different parameters in $G_1(z)$, $G_2(z)$, and $G_3(z)$. Recall that these three transfer functions together make up the full controller --- thus, the parameters for $G_1(z)$ and $G_3(z)$ will affect the numerical values of the $G_2(z)$ controller. By construction, all of these simulations correspond to the same controller output and closed-loop behavior; when combined with $G_1(z)$ and $G_3(z)$ (see Fig. \ref{fig:g1g3}), all circuits will produce the input-output characteristics shown in Fig. \ref{fig:cl_sims}. However, each circuit not only has a different shape, but a different set of neural firing patterns; furthermore, for a given circuit, different firing patterns can be produced from different realization matrices. For instance, in the full realization circuit (Fig. \ref{fig:full_diagram}), neuron $x_1$'s output can have a small change in firing rate (relative to equilibrium), or a large change in firing rate; additionally, the change can be positive or negative depending on the realization matrices. 

\section{Conclusions and future work} \label{sec:future_work}
In this work, we conducted a case study on a scalar neuromuscular system using linear neurons with self-dynamics. Given a sensorimotor system that behaves like an optimally-controlled closed-loop system, we can use methods from this work to provide plausible circuits and firing patterns that would implement the observed behavior. However, without additional information, we have no way to determine which of the circuits and firing patterns are the correct ones, i.e., what is the circuit that is actually found in the nervous system. A key contribution of this work is to provide methods to generate plausible circuits, and to highlight the fact that circuits (and even behaviors within the same circuit) are non-unique. To find out which circuit is the correct one, we require data from the relevant biological neuron(s). Conversely, if we have data from the biological neuron(s), we can tune the controller circuit to match this data while preserving the model's match to behavioral observations.

There are many directions of future investigation that would make this analysis more salient for neuroscience:

\begin{enumerate}
    \item Incorporate more realistic features of neurons (e.g., nonlinear dynamics, thresholds, spiking)
    \item Incorporate excitatory and inhibitory neurons. In animals, neurons are generally excitatory (i.e., all outputs have positive synaptic weights on downstream neurons) or inhibitory (i.e., all negative). This translates to sign requirements on internal variables in the controller realization and circuit structure        
    \item Use connectomic\footnote{The \textit{connectome} is a comprehensive map of neural connections in the nervous system. The connectome is fully mapped for some organisms (e.g. \textit{C. elegans}, fruit flies), and partially known for humans and other animals} information to build larger delay graphs and connectome-compatible controllers
\end{enumerate}

\bibliography{refs}
\bibliographystyle{IEEEtran}

\end{document}